\documentclass[article]{aastex631}
\usepackage{graphicx}   
\usepackage{amsmath}    

\usepackage{amssymb}    
\usepackage{multirow}

\received{?}
\revised{?}
\accepted{?}
\submitjournal{ApJL}
\shorttitle{Classification and Event Rate of $Swift$ GRBs}
\shortauthors{J. J. Luo et al.}
\graphicspath{{./}{figures/}}

\begin{document}

\title{The classification and
    formation rate of $\mathbf{Swift/BAT}$ gamma-ray bursts\footnote{Released on May, 3st, 2024}}

\author[0000-0002-1665-1268]{Juan-Juan Luo}
\affiliation{School of Physics and Electronics, Qiannan Normal University for Nationalities, Duyun 558000, P. R. China}\thanks{E-mail: j\_j\_luo@sina.com}

\author[0009-0006-6903-4454]{Liang Zhang}
\affiliation{Guizhou Vocational College of Economics and Business, Duyun 558022 , P. R. China}\thanks{E-mail: liang\_zhang\_gz@sina.com}

\author[0000-0002-2394-9521]{Li-Yun Zhang}
\affiliation{College of Science, Guizhou University and NAOC-GZU-Sponsored Center for Astronomy, Guizhou University, Guiyang 550025, P. R. China}\thanks{E-mail: liy\_zhang@hotmail.com}

\author[0000-0001-7199-2906]{Yong-Feng Huang}
\affiliation{School of Astronomy and Space Science, Nanjing University, Nanjing 210023, P. R. China}\thanks{E-mail: hyf@nju.edu.cn}
\affiliation{Key Laboratory of Modern Astronomy and Astrophysics (Nanjing University), Ministry of Education, P. R. China}

\author{Jia-Quan Lin}
\affiliation{Guizhou Vocational College of Economics and Business, Duyun 558022 , P. R. China}

\author{Jun-Wang Lu}
\affiliation{School of Physics and Electronics, Qiannan Normal University for Nationalities, Duyun 558000, P. R. China}

\author{Xiao-Fei Dong}
\affiliation{School of Astronomy and Space Science, Nanjing University, Nanjing 210023, P. R. China}



\begin{abstract}

Gamma-ray bursts (GRBs) are usually classified into
long/short categories according to their durations, but controversy
still exists in this aspect. Here we re-examine the long/short
classification of GRBs and further compare
the cosmological distribution and evolution of each potential
subclass. A large number of $Swift/BAT$ GRBs are analyzed
in this study. The Gaussian mixture model is used to fit the duration
distribution as well as the joint distribution of
duration and hardness ratio, and the Akaike and Bayesian information
criteria are adopted to assess the goodness of fit. It is found
that three Gaussian components can better fit both the univariate and
bivariate distributions, indicating that there are three subclasses
in the $Swift/BAT$ GRBs, namely short, intermediate,
and long subclasses. The non-parametric Efron-Petrosian and Lynden-Bell's
$c^{-}$ methods are used to derive the luminosity function and
formation rate from the truncated data of bursts with known redshift
in each subclass. It is found that the luminosity distributions
and birth rates of the three subclasses are different,
further supporting the existence of the
intermediate subclass in the $Swift/BAT$ GRBs.

\end{abstract}



\section{Introduction}
\label{sec:Introduction}

Thousands of Gamma-ray bursts (GRBs,
\cite{Klebesadel1973ApJ...182L..85K}) have been detected from the
sky up to now. Based on their durations ($T_{90}$,
\cite{1993ApJ...413L.101K}), GRBs can be generally be classified
into two groups, i.e. short GRBs (sGRBs) and long GRBs (lGRBs).
Short GRBs are believed to originate from mergers of double
neutron stars (NS-NS) or NS-black hole binaries, while lGRBs are
thought to come from the collapse of massive stars at the end of
their lives
\citep{1993ApJ...405..273W,1998ApJ...494L..45P,2006ARA&A..44..507W,2022MNRAS.517.5770Z}.
In this framework, the event rate of lGRBs should naturally trace
the star formation rate (SFR)
\citep{2010ApJ...711..495B,2012A&A...539A.113E,2013A&A...556A..90W}.

The idea of classifying GRBs according to their $T_{90}$ can be
traced back to the $CGRO/BATSE$ era. It was found that the
distribution of $T_{90}$ can be well fitted by two lognormal
components \citep{1993ApJ...413L.101K,1994MNRAS.271..662M}, which
correspond to sGRBs group and lGRBs group. Interestingly,
\cite{1998ApJ...508..757H} noticed that there is an intermediate
group located between short and long groups. However, as the
number of GRBs in the $CGRO/BATSE$ catalogue increases, the
existence of the intermediate component seems to be less apparent
\citep{2002A&A...392..791H,2015A&A...581A..29T,2022MNRAS.517.5770Z}.
Since then, such a third component has been widely paid attention
to and controversial results are obtained. For example,
\cite{2015Ap&SS.357....7Z} analyzed 248 $Swift/BAT$ GRBs with
known redshift, claiming that the intermediate group is present
rather clearly in the fitting of $T_{90}$ in both the observer and
rest frame. \cite{2015A&A...581A..29T} analyzed $\sim 1500$
$Fermi/GBM$ GRBs and also concluded that the $T_{90}$ distribution
is well fitted by a mixture of three log-normal distributions,
indicating the presence of the intermediate class. On the other
hand, \cite{2016MNRAS.462.3243Z} composed large samples consisting
of $2037$ $CGRO/BATSE$ GRBs, $956$ $Swift/BAT$ GRBs (298 events
with redshift available), and $1741$ $Fermi/GBM$ GRBs,
respectively. They found that the two-Gaussian (2-G) model is
strongly supported by $CGRO/BATSE$ and $Fermi/GBM$ samples, while
the $Swift/BAT$ sample supports the three-Gaussian (3-G) model.
Additionally, the number of GRBs in the sample has a significant
impact on the classification of GRBs. \cite{2022MNRAS.517.5770Z}
have conducted a detailed study on the impact of sample size on
the classification of GRBs and found that the intermediate group
could be concealed when the sample size is large.
Theoretically, an important reason that leads to the
doubtful existence of intermediate group is the lack of progenitor
stars that could produce these special events.

The $Swift/BAT$ catalogue includes many GRBs with redshift
available \citep{2004ApJ...611.1005G}, which makes it possible to
study the event rate of lGRBs. Based on these events, researchers
have investigated possible evolution of the luminosity function
with respect to redshift as well as the relationship between GRB
event rate and the SFR \citep{2007ApJ...662.1111L,
2010MNRAS.406.1944W,2011MNRAS.416.2174C,2012ApJ...749...68S,
2012ApJ...744...95R,2014MNRAS.444...15H,2015ApJ...806...44P,
2015ApJS..218...13Y,2016ApJ...820...66D,2016A&A...587A..40P,
2018ApJ...852....1Z,2021ApJ...914L..40D,2022MNRAS.513.1078D,
2024ApJ...963L..12P}. Therefore, studying the cosmological
distribution and evolution of each subclass of GRBs can provide
useful information on their physical origins.

In this study, we collect a large $Swift/BAT$ GRB sample,
based on which the univariate duration ($T_{90}$) distribution and
bivariate $T_{90}$ versus hardness ratio (HR) distribution are
analyzed. Various subclasses are identified and their redshift
distribution and cosmological evolution are compared. The
structure of our paper is organized as follows. A detailed
description of the data acquisition and the sample constitution is
presented in Section \ref{sec:DATASELECTION}. The methods adopted
to analyze the samples are described in Section \ref{sec:METHODs}.
Here, we use the Gaussian Mixture Model (GMM) to assess the
$T_{90}$ distribution and use both the Akaike information criteria
(AIC) and Bayesian information criteria (BIC) to evaluate the
fitting results comprehensively. Due to the unavoidable
observational selection effect of the $Swift/BAT$ detector, the
sample is a truncated and incomplete collection. We use the
non-parametric Efron-Petrosian (EP) method
\citep{1992ApJ...399..345E} to eliminate the redshift evolution of
the luminosity, and use the Lynden-Bell's $c^{-}$ method
\citep{1971MNRAS.155...95L} to derive the intrinsic luminosity
function and event rate of GRBs. Our numerical results are
presented in Section \ref{sec:Results}. Finally, Section
\ref{sec:CONCLUSION} presents our conclusions and discussion.
Throughout this study, we assume a flat $\Lambda$ CDM cosmology
with $\Omega_{\rm m} = 0.27$ and $H_0 = 70$ km s$^{-1}$
Mpc$^{-1}$.

\section{DATA SELECTION}
\label{sec:DATASELECTION}

As of February 18, 2024, the $Swift/BAT$ GRB Catalog
\footnote{https://swift.gsfc.nasa.gov/archive/grb\_table/}
contains 1832 GRBs, of which 440 events have redshift
measurements. In order to study the long/short
classification of GRBs, we have grouped the bursts into two
samples: Sample I contains all the 1512 $Swift/BAT$ GRBs with
valid $T_{90}$ data; Sample II contains 1454 $Swift/BAT$ GRBs with
both valid $T_{90}$ and spectral hardness ratio data. The hardness
ration is usually defined as
\begin{equation}\label{eq:eq1}
HR=\frac{S_{50-100keV}}{S_{15-25keV}}=\frac{\int^{100keV}_{50keV}F(E)EdE}{\int^{25keV}_{15keV}F(E)EdE},
\end{equation}
where $F(E)$ is the spectrum which could generally be fit by a
power-law or cut-off power-law function.

In order to study the cosmological distribution and
evolution of each subclass, the peak luminosity ($L_{\rm p})$ of
each GRB should be known, which further requires that several
other parameters should be available. They include the redshift,
the spectral index, as well as the peak flux in the energy range
of 15 --- 150 keV. In our study, $L_{\rm p}$ is calculated by
using $L_{\rm p} = 4 \pi {d_L}^2(z) F K(z)$, where $d_L(z)$
denotes the luminosity distance, $K(z)$ denotes the K-correction,
$F$ denotes the peak flux in units of photons cm$^{-2}$ s$^{-1}$
in the energy range of 15 keV
---  150 keV for $Swift/BAT$. \cite{2021ApJ...914L..40D}
pointed out that the luminosity evolution (i.e., the dependence of
luminosity on the redshift) is sensitive to the flux limit.
An improper selection of the flux limit could lead to a
false luminosity evolution. In our study, we adopt two flux
limits, $F_{\rm limit,1} = 2.0 \times 10^{-8}$ erg cm$^{-2}$
s$^{-1}$ and $F_{\rm limit,2} = 5.0 \times 10^{-9}$ erg cm$^{-2}$
s$^{-1}$, to overcome this problem as far as possible. $F_{\rm
limit,1}$ is used for short GRBs;
while $F_{\rm limit,2}$ is used for intermediate GRBs and long
events. The limiting luminosity is then
expressed as $L_{\rm limit} = 4 \pi {d_L}^2(z)F_{\rm limit}
\bar{K}(z)$ \citep{2015ApJ...806...44P}, where the K-correction
term $\bar{K}(z)$ is calculated by assuming a simple power law
spectrum with a mean spectral index of $-1.1$
\citep{2018PASP..130e4202Z}.

\section{methods}
\label{sec:METHODs} Fitting the $T_{90}$ distribution with
multiple Gaussian components is widely adopted in the long/short
classification of GRBs. The focus is how to optimize the objective
function to determine the appropriate number of Gaussian
components and obtain the parameters for each component. For
example, \cite{1998ApJ...508..757H,
2002A&A...392..791H,2015A&A...581A..29T,2015Ap&SS.357....7Z} used
$\chi^2$ fitting to analyze the $T_{90}$ distribution.
\cite{2014PASJ...66...42T,2016MNRAS.462.3243Z,2022MNRAS.517.5770Z}
used the Gaussian mixture model (GMM) algorithm to study the
long/short classification of GRBs. Since the $\chi^2$ fitting is
affected by input data binning strategy, here we adopt the GMM
algorithm in our study. The probability density function (PDF) of
$K$ components GMM is
\begin{equation}\label{eq:eq2}
p(\boldsymbol{x}) = \sum_{k=1}^{K}A_k
\mathcal{N}\left(\boldsymbol{x} ; \boldsymbol{\mu_k},\boldsymbol{\Sigma_k}\right),
\end{equation}
where $\mathcal{N}\left(\boldsymbol{x} ; \boldsymbol{\mu_k},\boldsymbol{\Sigma_k}\right)$ represents the PDF of
each component, $\boldsymbol{x}$ represents a M-dimensional data vector:
$\boldsymbol{x}=x_j(j=1,2,...,M)$ \citep{2014PASJ...66...42T}. The weights satisfy $\sum_{k=1}^{K} A_{k}=1$. For the $k$th Gaussian component, the PDF is
\begin{equation}\label{eq:eq3}
\mathcal{N} \left( \boldsymbol{x}; \boldsymbol{\mu_k}, \boldsymbol{\Sigma_k} \right) = \frac{1}{2\pi\sqrt{|\boldsymbol{\Sigma_k}|}} exp \left[-\frac{1}{2} \left(\boldsymbol{x} - \boldsymbol{\mu_k} \right)^{T} \boldsymbol{\Sigma_k}^{-1} \left( \boldsymbol{x} - \boldsymbol{\mu_k} \right) \right],
\end{equation}
where $\boldsymbol{\mu_i}$ is the mean value and $\boldsymbol{\Sigma_i}$ is the covariance
matrix, with $| \boldsymbol{\Sigma_i} | = det(\boldsymbol{\Sigma_i} )$.

\cite{2014PASJ...66...42T} pointed out that given a sequence of
independent data $\boldsymbol{X}
=\{\boldsymbol{x_1},\boldsymbol{x_2},...,\boldsymbol{x_N}\}$, the
logarithmic likelihood function is
\begin{equation}\label{eq:eq4}
\ln p\left( \boldsymbol{X}; \boldsymbol{A}, \boldsymbol{\mu},
\boldsymbol{\Sigma} \right) = \sum_{i=1}^{N}\ln
\left[\sum_{k=1}^{K} A_k\mathcal{N}\left(\boldsymbol{x_i} ;
\boldsymbol{\mu_k},\boldsymbol{\Sigma_k}\right) \right].
\end{equation}
The goal is to maximize the likelihood function to
determine the optimal parameters of the model. By taking partial
derivatives of the log likelihood function with respect to
$\boldsymbol{\mu_k}$ and $\boldsymbol{\Sigma_k}$ and letting them
equal to zero, expressions of the above parameters can be obtained
as
\begin{equation}\label{eq:eq5}
\boldsymbol{\mu_k} = \frac{1}{N_k}\sum_{i=1}^{N}\gamma_{ik}\boldsymbol{x_i},
\end{equation}
\begin{equation}\label{eq:eq6}
\boldsymbol{\Sigma_k} = \frac{1}{N_k}\sum_{i=1}^{N}\gamma_{ik}\left(\boldsymbol{x_i}-\boldsymbol{\mu_k} \right)\left(\boldsymbol{x_i}-\boldsymbol{\mu_k} \right)^T,
\end{equation}
where
\begin{equation}\label{eq:eq8}
N_k = \sum_{i=1}^{N}\gamma_{ik},
\end{equation}
and $\gamma_{ik}$ is called the responsibility, which represents
the probability that a particular event belongs to each cluster.
It can be calculated as
\begin{equation}\label{eq:eq7}
\gamma_{ik} = \frac{A_k\mathcal{N}\left(\boldsymbol{x_i} ;
\boldsymbol{\mu_k},\boldsymbol{\Sigma_k}\right)}{\sum_{j = 1}^{K}
A_j\mathcal{N}\left(\boldsymbol{x_i} ;
\boldsymbol{\mu_j},\boldsymbol{\Sigma_j}\right)}.
\end{equation}
The proportion coefficient $A_k$ of each Gaussian
component is determined by
\begin{equation}\label{eq:eq9}
A_k = \frac{N_k}{N},
\end{equation}
which is derived using the Lagrange multiplier method.

From Equations \ref{eq:eq5}, \ref{eq:eq6} and
\ref{eq:eq9}, it can be seen that the key to derive the model
parameters is to determine $\gamma_{ik}$. However, $\gamma_{ik}$
itself is a function of model parameters. Therefore, the
expectation-maximization algorithm is introduced in our study,
which is an iterative process used to solve the optimal parameters
of probability models containing latent variables
\citep{2014PASJ...66...42T}. The steps are as following: (i)
Initialize the model parameters of $A_k$, $\boldsymbol{\mu_k}$,
$\boldsymbol{\Sigma_k}$, and calculate the value of the
logarithmic likelihood; (ii) Calculate the $\gamma_{ik}$ values
using Equation \ref{eq:eq7}; (iii) Recalculate the values of the
model parameters using $\gamma_{ik}$ obtained in Step ii; (iv)
Repeat Step (ii) and Step (iii) until the likelihood function
converge. In this way, the optimal model parameters can be
self-consistently determined. Note that the GMM and EM algorithm
has been implemented in the \textbf{PYTHON} machine learning
package of \textbf{sklearn.mixture}
\footnote{https://scikit-learn.org/stable/modules/mixture.html\#gaussian-mixture},
which is used in our study.

The Akaike information criterion (AIC, \cite{1974ITAC...19..716A})
and Bayesian information criterion (BIC,
\cite{10.1214/aos/1176344136}) are widely used in assessing the
goodness of models in astronomy \citep{2007MNRAS.377L..74L}. They
are defined as $AIC = 2p - 2 \ln P_{\rm max}$ and $BIC = p\ln N -
2 \ln P_{\rm max}$, respectively, where $p$ represents the number
of parameters of the model, $N$ represents the sample size, and
$P_{\rm max}$ represents the maximum value of the likelihood
function. It is worth noting that when $N>8$, the first term of
AIC and BIC satisfies $p \ln N > 2p$, indicating that the penalty
term of BIC is much larger than that of AIC, especially for a
large sample size. Furthermore, a larger penalty term in BIC may
lead to underfitting (resulting in a simple model with fewer
parameters), while a smaller penalty term in AIC may lead to
overfitting (resulting in a complex model with more parameters).
These criteria for model selection have been widely
applied in previous studies on GRB classification
\citep{2014PASJ...66...42T,
2016MNRAS.462.3243Z,2019ApJ...870..105T,2022MNRAS.517.5770Z}.
Therefore, for the current GRB long/short classification problem,
we adopt the AIC and BIC criteria to determine the best fitting
model for each sample, which can achieve a good balance between
the two competing terms so that it generally has the minimum AIC
and BIC values (hereafter $AIC_{\rm min}$, $BIC_{\rm min}$). For
the $i$th model, the AIC difference is calculated as $\Delta
AIC_i=AIC_i-AIC_{\rm min}$, and the BIC difference is $\Delta
BIC_i=BIC_i-BIC_{\rm min}$. The range of $\Delta AIC_i$ and
$\Delta BIC_i$ and the corresponding descriptions are detailedly
shown in Table \ref{tab:table1}.

\begin{table*}
\renewcommand{\thetable}{\arabic{table}}
\centering \scriptsize \tabcolsep 0.5truecm \caption{The ranges
and corresponding descriptions of $\Delta AIC_i$ and $\Delta
BIC_i$}\label{tab:table1}
\begin{tabular}{ccc}
\hline
\hline
name          & range                 &  description$^{a}$   \\
\hline

\multirow{4}{*}{$\Delta AIC_i$}   & (0,2) & a substantial support for the $i$th model       \\
                                 &  (2,4) &a strong support for the $i$th model             \\
                                 &  (4,7) &a considerably less support for the $i$th model  \\
                                 &  $>$7  & essentially no support for the $i$th model     \\

\multirow{4}{*}{$\Delta BIC_i$}   & (0,2) & a substantial support for the $i$th model \\
                                 &  (2,6) & an evidence against the $i$th model  \\
                                 &  (6,10)& a strong evidence against the $i$th model  \\
                                 &  $>$10 &  a very strong evidence against the $i$th model   \\
\hline
\end{tabular}

{$\textbf{Note}$. $^{a}$References:
\cite{doi:10.1080/01621459.1995.10476572,doi:10.1177/0049124104268644,2019ApJ...870..105T,2022MNRAS.517.5770Z}}

\end{table*}

Previous studies indicate that the luminosity ($L$) and
redshift ($z$) of GRBs are not independent. Determination of the
cosmological GRB distribution requires that the  $L$ -- $z$
relation should be established first. \cite{1992ApJ...399..345E}
developed a useful method that is very effective in this aspect.
It can be applied to both one sided truncation
\citep{1992ApJ...399..345E} and two sided truncation samples
\citep{efron1998}. The Efron-Petrosian (EP) method has been
adopted to determine the luminosity evolution in many astronomical
fields \citep{2004ApJ...609..935Y,
2015ApJS..218...13Y,2017ApJ...850..161T,
2018ApJ...852....1Z,2022MNRAS.513.1078D}. In this study, we also
use the EP method to derive the luminosity evolution of GRBs,
which is characterized by a redshift-dependent term of $g(z)$. The
de-evolved luminosity in the GRB local frame can then be
calculated as $L_0 = L / g(z)$. As usual, we assume that $g(z)$
takes the form of a power-law function, $g(z) = (1+z)^k$. Details
on determining the index $k$ through the EP method can be found in
\cite{1992ApJ...399..345E}.

After deriving the de-evolved luminosity $L_0$, the local
cumulative luminosity function $\psi(L_{0})$ and the cumulative
redshift distribution $\phi(z)$ can be derived by using the
Lynden-Bell's $c^{-}$ method \citep{1971MNRAS.155...95L}. This
method was rediscovered for its wide application in recovering
truncated luminosity functions by \citep{10.1214/aos/1176346584}
and \cite{10.1214/aos/1176350180}, and is extensively discussed in
\cite{1988MNRAS.232..431E,1992ApJ...399..345E,Dorre2019}. It is a
unique nonparametric procedure applicable to randomly truncated
univariate data, which has already been interestingly used in the
field of GRBs (e.g. \cite{2013ApJ...774..157D,
2015MNRAS.451.3898D,2017ApJ...848...88D,2017A&A...600A..98D,
2020ApJ...904...97D,2021ApJ...914L..40D,2021Galax...9...95D,
2022MNRAS.514.1828D,1999ApJ...511..550L,2004ApJ...609..935Y,
2015ApJS..218...13Y,2017ApJ...850..161T,2018ApJ...852....1Z,
2022MNRAS.513.1078D}). It is worth noting that our samples are
also truncated data since the energy range and sensitivity (the
flux threshold) of $Swift/BAT$ are always limited during the
observations. Considering these unavoidable observational
selection effects, we will use the non-parametric Efron-Petrosian
and Lynden-Bell's $c^{-}$  method (EP-L, for short) to study the
cosmological distribution and evolution of different subclasses of
GRBs. For details on determining the local cumulative luminosity
function $\psi(L_{0})$ and the cumulative redshift distribution
$\phi(z)$ of $Swift/BAT$ GRBs through the EP-L method, please refer
to \cite{2015ApJ...806...44P,
2015ApJS..218...13Y,2022MNRAS.513.1078D}.

\section{Results} \label{sec:Results}

\subsection{Long/short classification of GRBs}
\label{sec:Long/short classification of GRBs}

\begin{table*}
\scriptsize
\centering
\tabcolsep 0.2truecm
\caption{The results
of AIC and BIC values for Samples I and II. For the
distribution of GRBs in each sample, a mixture of 1 to 6 Gaussian
components are tested. }
\label{tab:table2}
\begin{tabular}{ccccc}
\hline
\hline

\multirow{2}{*}{No.$^{a}$}
& \multicolumn{2}{c}{  Sample I$^{c}$} & \multicolumn{2}{c}{ Sample II$^{c}$ }\\
& AIC & BIC & AIC & BIC\\
\hline
min$^{b}$ & $\mathbf{3265.21}$ & $\mathbf{3307.78}$ &
$\mathbf{2675.07}$ & $\mathbf{2764.87}$
\\

$1  $ & $333.65 $ & $301.72 $ & $506.45 $ & $443.07 $\\

$2  $ & $30.88  $ & $14.92  $ & $75.45  $ & $43.76  $ \\

$3  $ & $0       $ & $0     $ & $0      $ & $0      $ \\

$4  $ & $12.18  $ & $28.15  $ & $1.38  $ & $33.07  $  \\

$5  $ & $16.50  $ & $48.43  $ & $90.3  $ & $72.41  $  \\

$6  $ & $17.19  $ & $65.08  $ & $15.52  $ & $110.60  $ \\

\hline

\end{tabular}

{$\textbf{Note}$. $^a$The number of Gaussian component.
       $^b$The ``min'' row lists the minimum AIC and BIC values
       obtained for each sample with GMM
       of 1 to 6 components, which correspond to the
       best fit model. The data marked in boldface in this row are true AIC
       and BIC values. The data in other rows are differential values of $\Delta AIC$
       and $\Delta BIC$ with respect to the ``min'' row.
       $^c$The number of GRBs in Samples I and II is $1512$, $1401$, respectively.}

\end{table*}

\begin{figure}[ht!]
\gridline{\fig{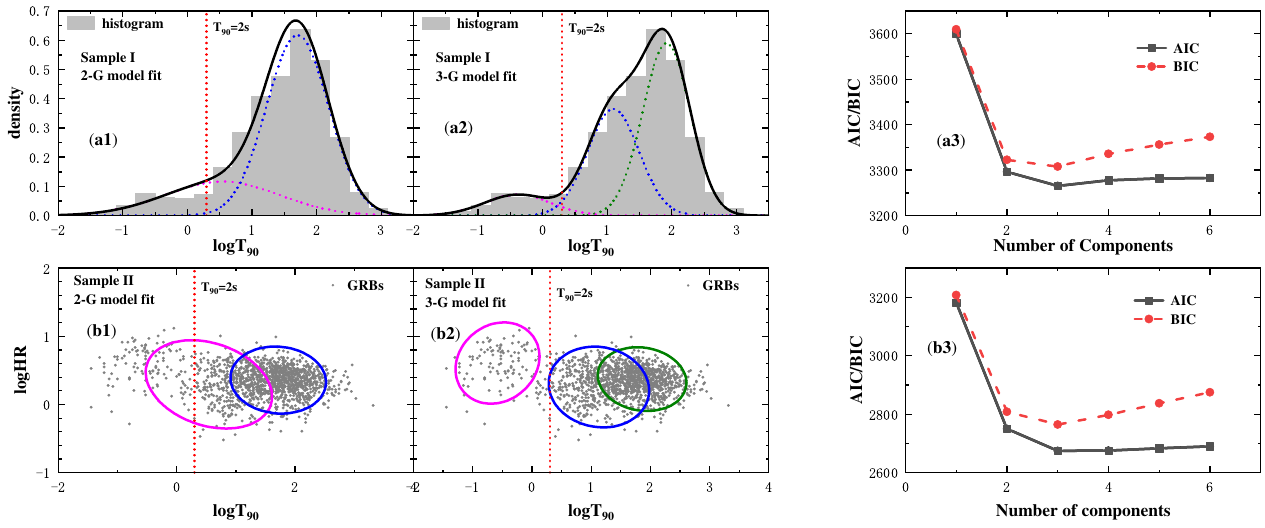}{1\textwidth}{}}
 \caption{Fitting Samples I and II by using the GMM models.
         The upper panels show the $T_{90}$ distribution of Sample I
         fitted with the 2-G and 3-G models, together with the AIC/BIC values. In
         (a1) and (a2) panels, the solid curves show the superposed total PDF of
         the 2-G model and 3-G model, while the dashed curves show the contribution
         of different components. The lower panels illustrate the joint distribution of
         $T_{90}$ -- HR for Sample II, fitted with the 2-G and 3-G models.
         Both the X- and Y-axis are in logarithmic coordinates. The (b1) panel shows the
         full width of half maximum (FWHM) of
         each component in the framework of the 2-G model, and the (b2) panel
         correspondingly shows the 3-G model fitting results. The dotted vertical lines
         represent $T_{90} = 2$ s. In (a3) and (b3) panels, it could be seen that both AIC and
         BIC criteria support the 3-G model for Samples I and II.
         \label{fig:1_figure}}
\end{figure}

We have used the one-dimensional GMM model to fit the
$T_{90}$ distribution of Sample I. A two-dimensional GMM model is
also applied on Sample II to consider the joint effects of
$T_{90}$ and HR on the classification. The AIC and BIC values
obtained in the fitting are presented in Table \ref{tab:table2}.
Figure \ref{fig:1_figure} illustrates the fitting results of the
2-G model and 3-G model to these two samples.

From the second and third columns of Table
\ref{tab:table2}, it can be seen that the 3-component model gives
the best description for the $T_{90}$ distribution of Sample I,
which is consistent with previous
 studies \citep{2002A&A...392..791H,2015A&A...581A..29T,2019ApJ...870..105T,2016MNRAS.462.3243Z}.
The minimum values of AIC and BIC are correspondingly $AIC_{3} =
3265.21$ and $BIC_{3}= 3307.78$. Additionally, as a comparison, we
have $\Delta AIC_{2}= 30.88$ and $\Delta BIC_{2}= 14.92$ for the
two component model. \cite{2016MNRAS.462.3243Z} have conducted a
similar study using two $Swift/BAT$ samples (with 708 and 956 GRBs
respectively). They concluded that the 3-G model was the best
fitting model. In their work, $\Delta BIC_{2}$ is approximately
5.5 for the two component model. Considering that the number of
GRBs in our Sample I is more than twice of that of
\cite{2016MNRAS.462.3243Z}, a larger $\Delta BIC_{2}$ obtained by
us thus can be regarded as a strong support for the existence of
the intermediate subclass. In the upper panels of Figure
\ref{fig:1_figure}, we could also see that the 3-G model generally
presents a better fit to the observations than the 2-G model.
Panel (a3) shows that both AIC and BIC tests support the 3-G model
for Sample I.

Similarly, the fourth and fifth columns of Table
\ref{tab:table2} show that the 3-component model also gives the
best description for the $T_{90}$ -- HR distribution of Sample II.
The minimum values of AIC and BIC are correspondingly $AIC_{3} =
2675.07$ and $BIC_{3}= 2764.87$, which is consistent with the
above conclusion drawn by only considering the $T_{90}$ parameter.
In the lower panels of Figure \ref{fig:1_figure}, there are two
contour curves in the (b1) panel and three contour curves in the
(b2) panel. The pink contour mainly includes sGRBs while the green
contour contains most of the lGRBs. The blue contour in the (b2)
panel corresponds to the intermediate component, which are also
relatively long GRBs. We could see that the 3-G model generally
presents a better fit to the observations than the 2-G model.
Especially, we notice that for the 2-G model, we have $\Delta
AIC_{2}= 75.45$ and $\Delta BIC_{2}= 43.76$ in Sample II, which
are correspondingly larger than that in Sample I ($\Delta AIC_{2}=
30.88$ and $\Delta BIC_{2}= 14.92$). So, the support for the 3-G
model is further strengthened in Sample II as compared with Sample
I. In short, the inclusion of the HR parameter further strengthens
the existence of the intermediate subclass of GRBs.

\subsection{Luminosity function and event rate}
\label{sec:Luminosity function and event rate}

\begin{figure}[ht!]
\gridline{\fig{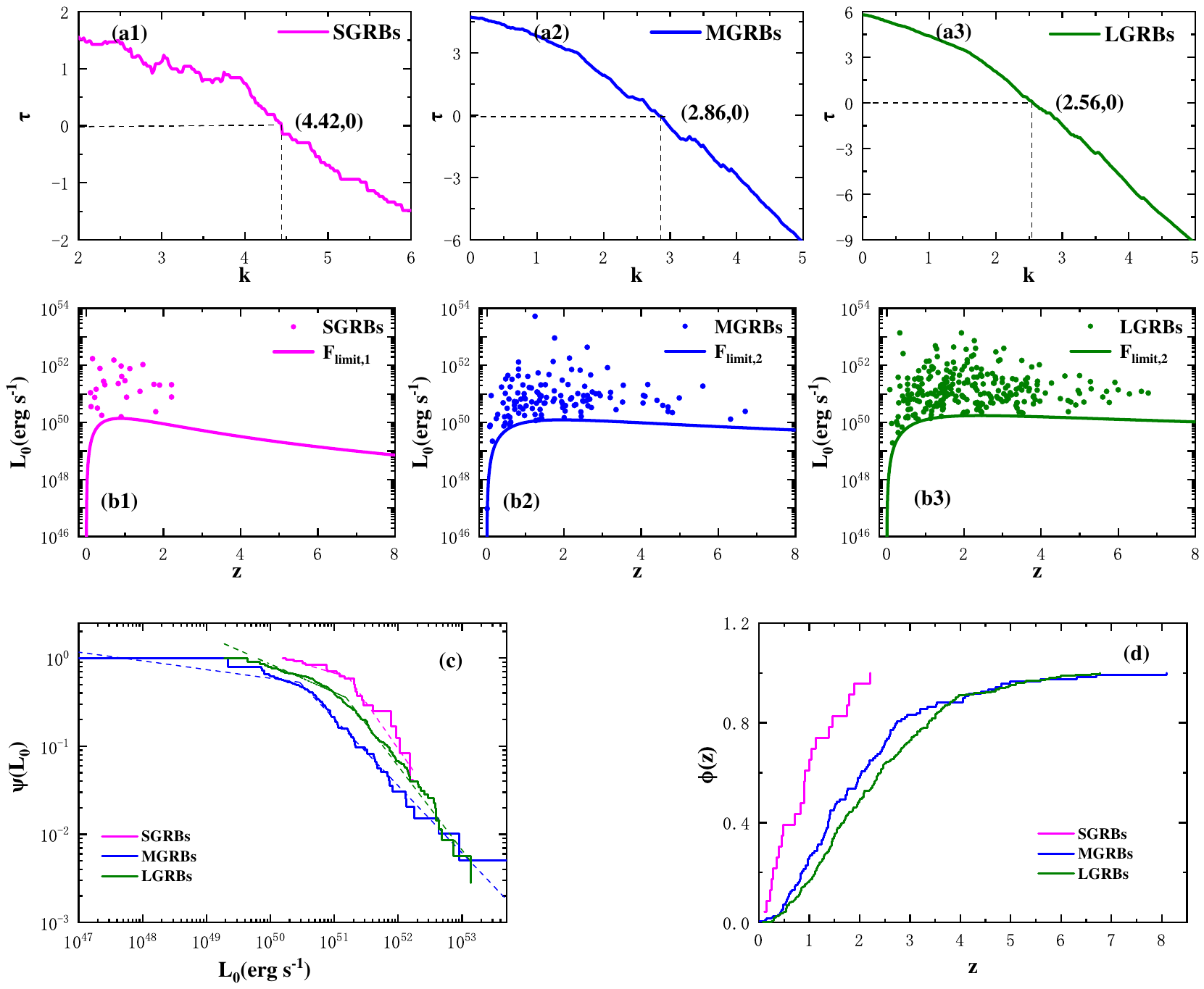}{1.0\textwidth}{}}
 \caption{Different features of the three subclasses of GRBs.
   Panels (a1), (a2), and (a3) plot the test statistic
   $\tau$ versus $k$ for SGRBs, MGRBs, and LGRBs, respectively.
   Panels (b1), (b2), and (b3) plot the de-evolved luminosity
   ($L_{0}$) versus the redshift ($z$) for the three samples. The
   solid lines represent the lower luminosity limits calculated from the two flux
   limits of $F_{\rm limit,1}$ and $F_{\rm limit,2}$, where
   $F_{\rm limit,1}$ is adopted for SGRBs,
   and $F_{\rm limit,2}$ is adopted for MGRBs and LGRBs.
   Panel (c) shows the cumulative luminosity function $\psi(L_{0})$
   of three samples, which is normalized to unity at
   the lowest luminosity. The dashed lines illustrate the best fit to
   the luminosity function by using a broken power-law function.
   Panel (d) shows the normalized cumulative redshift distribution of
   the three samples.
\label{fig:2_figure}
}
\end{figure}

According to the above analysis, GRBs should be classified into
three groups, the short group (SGRBs), intermediate group (MGRBs),
and long group (LGRBs). Here we go further to study the
cosmological distribution and evolution of these three groups
using the EP-L method. For this purpose, an additional parameter
of redshift ($z$) is required to calculate the peak luminosity
$L_{\rm p}$ of each GRB (see Section \ref{sec:DATASELECTION}).
Ultimately, for SGRBs, there are only 25 GRBs meeting this
requirement. Similarly, for MGRBs and LGRBs , the numbers of
available events are 126 and 255, respectively. In our analysis
below, the flux limit $F_{\rm limit,1}$ is adopted for SGRBs,
while $F_{\rm limit,2}$ is adopted for MGRBs and LGRBs.

As the first step, we use the non-parametric
Efron-Petrosian (EP) method to determine the luminosity evolution
of each group. As described in Section \ref{sec:METHODs}, the
luminosity evolution is assumed to take the form of $g(z) =
(1+z)^k$. The panels of (a1), (a2), and (a3) in Figure
\ref{fig:2_figure} illustrate the test statistic $\tau$ versus $k$
for the three subclasses of GRBs. For SGRBs, the $k$ index is
derived as $k = 4.43^{+1.03}_{-1.06}$ at $1 \sigma$ confidence
level. For MGRBs and LGRBs, the indices are $k =
2.86^{+0.43}_{-0.55}$ and $k = 2.56^{+0.33}_{-0.27}$,
respectively. Obviously, the $k$ parameter is different for the
three groups. Our $k$ value of SGRBs is consistent with previous
works that studied the luminosity evolution of short GRBs
\citep{2018ApJ...852....1Z,2021ApJ...914L..40D}, and our $k$ value
of LGRBs is also consistent with previous results concerning long
GRBs
\citep{2004ApJ...609..935Y,2015ApJS..218...13Y,2015ApJ...806...44P,
2016A&A...587A..40P}. Using the $k$ parameter, the de-evolved
luminosity can be calculated as $L_{0} = L / (1 + z)^{4.43}$,
$L_{0} = L / (1 + z)^{2.86}$, and $L_{0} = L / (1 + z)^{2.56}$ for
SGRBs, MGRBs and LGRBs, respectively. The de-evolved luminosity
$L_{0}$ versus redshift $z$ of the three samples are illustrated
in the panels of (b1), (b2), and (b3) of Figure
\ref{fig:2_figure}.

Then, we can use the Lynden-Bell's $c^{-}$ method to
derive the non-evolving cumulative luminosity function
$\psi(L_{0})$ and the  cumulative redshift distribution $\phi(z)$
of SGRBs, MGRBs, and LGRBs. The results are shown in Panels (c)
and (d) of Figure \ref{fig:2_figure}, respectively. From Panel
(c), it can be seen that the profiles of $\psi(L_{0})$ are
different from for the three groups. Especially, the $\psi(L_{0})$
of MGRBs is lower than those of LGRBs and SGRBs, indicating that
MGRBs are mainly low luminosity bursts. The Kolmogorov-Smirnov
(K-S) test has been adopted to compare the cumulative luminosity
function of each pair of the three groups, which gives the
P-values as $p_{12} = 7.9 \times 10^{-3}$, $p_{13} = 4.3 \times 10^{-2}$, and
$p_{23} = 0.21$. Again, it strongly indicates that
the intermediate class does exist, rather than being a vassal of
the long class. A broken power-law function has been adopted to
fit the normalized cumulative luminosity function. For SGRBs, we
get the best fit result as
\begin{equation}\label{eq:eq10}
\psi(L_{0}) \propto \left \{
\begin{aligned}
L_{0}^{-0.22\pm0.01}, L_{0} \leq 1.68 \times 10^{51} {\rm erg}, \\
L_{0}^{-1.04\pm0.01}, L_{0} > 1.68 \times 10^{51} {\rm erg}.
\end{aligned}
\right.
\end{equation}
For MGRBs, we have
\begin{equation}\label{eq:eq11}
\psi(L_{0}) \propto \left \{
\begin{aligned}
L_{0}^{-0.10\pm0.01}, L_{0} \leq 2.89 \times 10^{50} {\rm erg}, \\
L_{0}^{-0.76\pm0.01}, L_{0} > 2.89 \times 10^{50} {\rm erg}.
\end{aligned}
\right.
\end{equation}
For LGRBs, we have
\begin{equation}\label{eq:eq12}
\psi(L_{0}) \propto \left \{
\begin{aligned}
L_{0}^{-0.32\pm0.01}, L_{0} \leq 1.52 \times 10^{51} {\rm erg}, \\
L_{0}^{-0.95\pm0.01}, L_{0} > 1.52 \times 10^{51} {\rm erg}.
\end{aligned}
\right.
\end{equation}
We see that the best fit functions are different for the
three groups, both in the characteristic breaking energy and the
power-law indices.

\begin{figure}[ht!]
\gridline{\fig{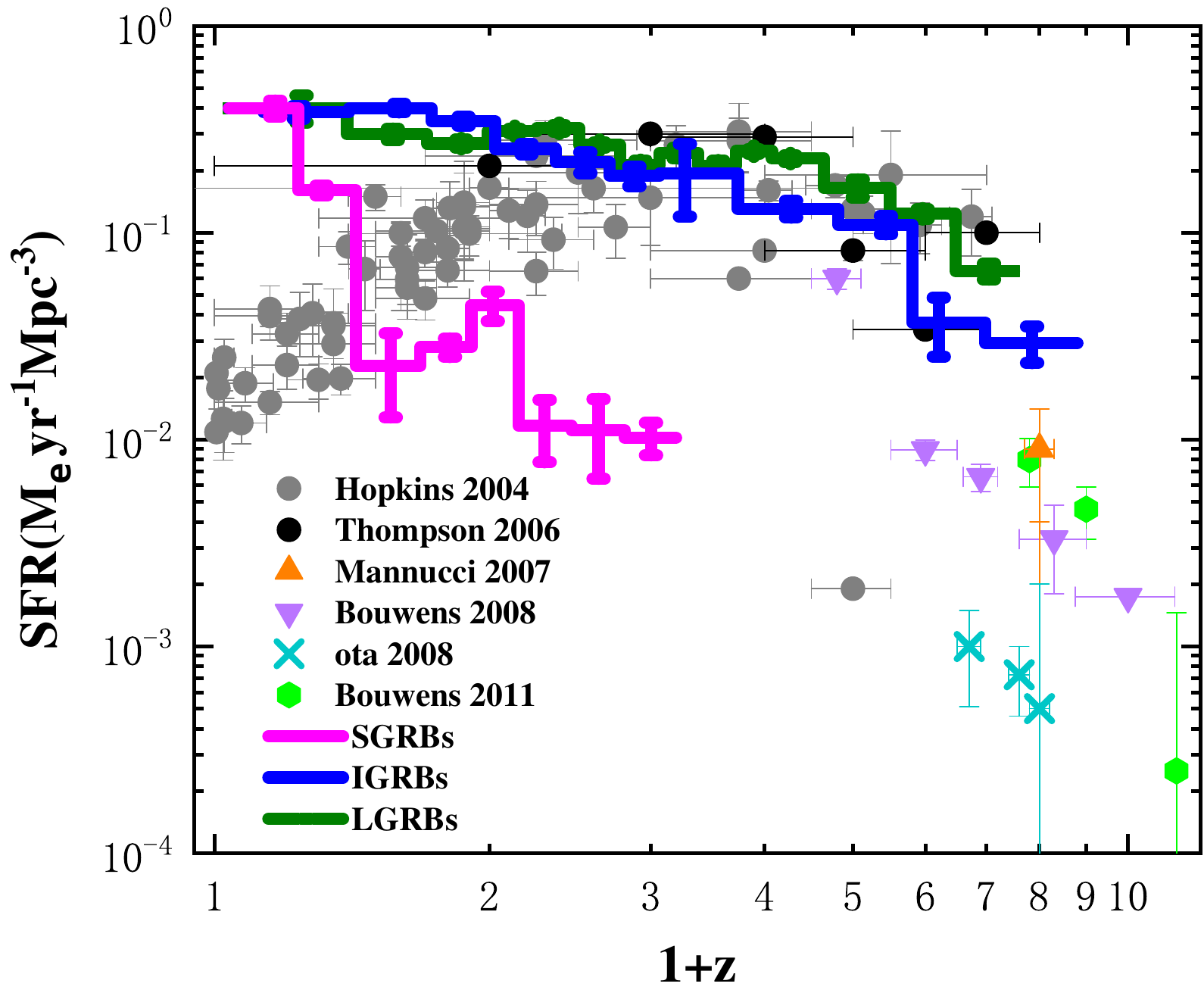}{0.4\textwidth}{}}
 \caption{The formation rate of GRBs ($\rho(z)$) versus redshift for
   SGRBs (heavy pink stepped line), MGRBs (blue stepped line), and
   LGRBs (green stepped line). The rate is normalized to unity at
   the first data point. A simple power-law function is
   adopted to fit $\rho(z)$.
   The resultant slopes are $-3.58 \pm 0.61$, $-1.54 \pm 0.19$, and $-0.79 \pm 0.12$,
   for SGRBs, MGRBs, and LGRBs, respectively.
   For comparison, the observed SFR are also plotted, with the data
   taken from \protect\cite{2004ApJ...615..209H} (gray dots),
   \protect\cite{2006ApJ...647..787T} (black dots),
   \protect\cite{2007A&A...461..423M} (triangles),
   \protect\cite{2008ApJ...677...12O} (crosses),
   \protect\cite{2008ApJ...686..230B} (inverted triangles),
   \protect\cite{2011Natur.469..504B} (orange pentagons).
   \label{fig:3_figure}}
\end{figure}

Panel (d) of Figure \ref{fig:2_figure} shows the
cumulative redshift distribution $\phi(z)$ of SGRBs, MGRBs, and
LGRBs, correspondingly. We see that
$\phi(z)$ are different for the three
subclasses. The formation rate ($\rho(z)$) of GRBs can be further
calculated from $d\phi(z)/dz$ as
\begin{equation}
 \label{eq:eq13}
\rho(z) = \frac{d\phi(z)}{dz}(1+z)\left(\frac{dV(z)}{dz}\right)^{-1},
\end{equation}
where
\begin{equation}
 \label{eq:eq14}
\frac{dV(z)}{dz}  = \frac{c}{H_{0}} \frac{4 \pi
d_{l}^{2}(z)}{(1+z)^{2}} \frac{1}{\sqrt{1- \Omega_m +
\Omega_m(1+z)^3}}.
\end{equation}
Figure \ref{fig:3_figure} illustrates the GRB formation
rate as a function of redshift, plotted for SGRBs, MGRBs, and
LGRBs, respectively. For comparison, the observed SFR data taken
from \cite{2004ApJ...615..209H}, \cite{2006ApJ...647..787T},
\cite{2007A&A...461..423M}, \cite{2008ApJ...677...12O},
\cite{2008ApJ...686..230B}, \cite{2011Natur.469..504B} are also
plotted. Generally, the formation rates of the three
subclasses decrease monotonously as the redshift increases. They
all clearly exceed the SFR at $z < 1$. Additionally, the formation
rates are obviously different for the three groups. Especially,
the $\rho(z)$ of SGRBs differs from that of the other two groups
significantly. A simple power-law function has been adopted to fit
the $\rho(z)$ of these three samples. The power-law indices are
$-3.58\pm0.61$, $-1.54\pm0.19$, and $-0.79\pm0.12$ for SGRBs,
MGRBs, and LGRBs, respectively. The K-S test has been adopted to
compare the formation rate of each pair of the three groups, which
gives the P-values as $p_{12} = 6.1 \times 10^{-2}$, $p_{13} = 2.7
\times 10^{-3}$, and $p_{23} = 0.43$. It further supports that the
the three subgroups are distinct categories.

\section{CONCLUSIONS}
\label{sec:CONCLUSION}

In this study, we investigate the long/short classification of
$Swift/BAT$ GRBs and the corresponding redshift distribution and
event rate of each subclass. The intrinsic characteristic spectral
hardness ratio is further included as a useful supplement for the
classification. The GMM models are used to fit the $T_{90}$
distribution of GRBs, and both AIC and BIC criteria are adopted to
assess the goodness of fit. It is found that the $T_{90}$
distribution can be well fitted by the 3-G model, which means they
can be grouped into three classes, i.e. short GRBs, intermediate
GRBs, and long GRBs. Comparing with the results of
\cite{2016MNRAS.462.3243Z}, it is also found that the presence of
intermediate class becomes more significant as the number of GRBs
in the sample increases. When the spectral hardness ratio is
included, the best fitting model is still 3-G. For GRBs with a
known redshift so that the peak luminosity is available, we go
further to study the redshift distribution and event rate of the
three subclasses (SGRBs, MGRBS, and LGRBs) by using the EP-L
method. It is found that the luminosity evolves with
redshift as $L = L_{0}(1+z)^k$, with $k$ = 4.43, 2.86, 2.56 for
SGRBs, MGRBs and LGRBs, respectively. The results are consistent
with previous studies in the cases of short GRBs
\citep{2018ApJ...852....1Z,2021ApJ...914L..40D}, and long GRBs
\citep{2004ApJ...609..935Y,2015ApJS..218...13Y,2015ApJ...806...44P,
2016A&A...587A..40P}. Interestingly, for the three subclasses,
significant difference exists in the cumulative luminosity
function $\psi(L_{0})$ as well as in the event rate $\rho(z)$,
further supporting the existence of the intermediate subclass.

For the intermediate subclass GRBs, we notice that their
$T_{90}$ are mostly larger than 2s but less than 30s, and their
spectra are relatively soft, which make them quite similar to
classical long GRBs. However, the synthesized information from the
redshift distribution and event rate indicates that they form a
distinct subclass. Note that observational selection effect may
still exist and could lead to some confusion in the classification
of GRBs \citep{2013ApJ...763...15Q}. A significantly expanded
sample size will help overcome this problem in the future. On the
Other hand, the existence of the intermediate subclass of GRBs is
not completely unexpected in theoretical aspect. In fact, it is
not necessary that all long bursts with $T_{90}>2s$ should
uniquely originate from the collapse of massive stars. For
example, it has been argued that some events may be due to the
kick of high speed neutron stars \citep{2003ApJ...594..919H,
2022MNRAS.509.4916X}. In addition, the existence of GRBs with
extended emission component is another factor that challenges the
traditional long/short GRB classification. It is expected that
mechanisms other than traditional collapsars and binary compact
star mergers may produce at least a small portion of GRBs.

\section*{Acknowledgements}

\begin{acknowledgments}

We thank the anonymous referee for valuable suggestions that led
to an overall improvement of this study. Our work was supported
 by the National Natural Science Foundation of China (Grant Nos. 12233002, 12373032, 12365011),
 by the National Key R\&D Program of China (2021YFA0718500),
 and by the National SKA Program of China No. 2020SKA0120300,
 by the Youth Science \& Technology Talents Development Project
 of Guizhou Education Department (No. KY[2022]098), by the Project
 of Guizhou Vocational College of Economics and Business (No. Gzjm[2023]01).
 YFH also acknowledges the support from the Xinjiang Tianchi Program.

\end{acknowledgments}

%





\bibliography{example1}{}
\bibliographystyle{aasjournal}



\end{document}